\pdfoutput=1
\documentclass[a4paper,fleqn,usenatbib]{mnras}

\usepackage{txfonts}

\usepackage[T1]{fontenc}
\usepackage{ae,aecompl}

\usepackage{graphicx}	
\usepackage{amssymb}	

\title{Merging Binaries in the Galactic Center: 
The eccentric Kozai-Lidov mechanism with stellar evolution}

\author[A. P. Stephan et al.]{
Alexander P. Stephan,$^1$\thanks{E-mail: alexpstephan@astro.ucla.edu}
Smadar Naoz,$^1$
Andrea M. Ghez,$^1$ 
Gunther Witzel,$^1$
\newauthor{
Breann N. Sitarski,$^1$
Tuan Do$^1$
and Bence Kocsis$^2$}
\\
$^{1}$Department of Physics and Astronomy, University of California, Los Angeles, CA 90095, USA\\
$^{2}$Institute of Physics, E\"otv\"os University, P\'azm\'any P. s. 1/A, Budapest, 1117, Hungary
}

\date{Accepted XXX. Received YYY; in original form ZZZ}

\pubyear{2016}

\begin{document}
\label{firstpage}
\pagerange{\pageref{firstpage}--\pageref{lastpage}}
\maketitle

\begin{abstract}
Most, if not all, stars in the field are born in binary configurations or higher multiplicity systems. In dense stellar environment such as the Galactic Center (GC), many stars are expected to be in binary configurations as well. These binaries form hierarchical triple body systems, with the massive black hole (MBH) as the third, distant object. The stellar binaries are expected to undergo large amplitude eccentricity and inclination oscillations via the so-called ``eccentric Kozai-Lidov" (EKL) mechanism. These eccentricity excitations, combined with post main sequence stellar evolution, can drive the inner stellar binaries to merge. We study the mergers of stellar binaries in the inner $0.1$~pc of the GC caused by gravitational perturbations due to the MBH. We run a large set of Monte Carlo simulations that include the secular evolution of the orbits, general relativistic precession, tides, and post-main-sequence stellar evolution. We find that about $13 \%$ of the initial binary population will have merged after a few million years and about $29 \%$ after a few billion years. These expected merged systems represent a new class of objects at the GC and we speculate that they are connected to G2-like objects and the young stellar population.
\end{abstract}




\section{Introduction}\label{sec:Intro}

The proximity of the Galactic Center (GC) provides an accessible laboratory for studying different physical processes in the presence of a massive black hole (MBH), many of which may also take place in many other galactic nuclei. Observations of the GC give an exquisite opportunity to test different theoretical arguments and physical processes that involve MBHs and dense environments. Binary populations within the central 1 pc play a significant role in numerous processes that take place at the GC, including the relaxation state of the GC \citep{AlexHopman2009}\footnote{If the GC is relaxed, a dense, mass-segregated cusp of stellar mass BHs is expected near the MBH.}, the stellar number density \citep[e.g.,][]{AlexPfuhl2014,Prodan+2015}, the S-star cluster population \citep[e.g.,][]{Antonini+2013}, as well as hypervelocity stars \citep[e.g.,][]{Hills1988,YuTremaine2003,Ginsberg+2007,Peretsetal2009,Perets09}. Furthermore, compact object binaries in the GC are a potential source of gravitational wave (GW) emission \citep[e.g.,][]{OLeary2009,AntPer2012}.

Currently there are three confirmed observed  binaries in the inner $\sim0.2$~pc of the GC. The first  confirmed binary, IRS 16SW, is an equal mass binary ($m_{primary} = m_{secondary} = 50$~$M_\odot$) at a projected distance estimated as  $\sim0.05$~pc from the MBH with a period of  $19.5$~days \citep{Ott+1999,Martins+06}. Recently, \citet{Pfuhl+13} discovered  two additional binaries, an eclipsing Wolf-Rayet binary with a period of $2.3$~days,
and a long-period binary with an eccentricity of $0.3$ and a period of $224$~days. Both of these binaries are estimated to be at only $\sim0.1$~pc from the  MBH. The long-period binary detection provides lower limits on the $2$-body relaxation timescale, and an upper limit on the number density of the faint stars and the compact remnants (i.e., the dark cusp) that are expected to exist near the MBH, see \citet{Bahcall+77,AlexHopman2009,AlexPfuhl2014}.  The latter study is extremely interesting as it lays out the dynamical consequences of even a single detection of a long period binary. This stresses the need for more observations and
that combining them with the understanding of the dynamics will allow us to draw better conclusions and tighter constraints on the binary fraction. Recent observational endeavors were proven to be very promising in placing some limits on the binary fraction and the dynamical state of the GC.
These suggest that the total massive binary fraction in the GC is comparable to the galactic binary fraction \citep[e.g.,][]{Ott+1999, Rafelski+2007}. Massive O-star binaries at the GC are estimated to be about $7\%$ of the total massive stellar population \citep{Rafelski+2007}. These binaries (and those outside the central parsec) are mostly observed as eclipsing or ellipsoidal variable binaries \citep[e.g.,][]{Ott+1999, Rafelski+2007}. Lower mass binaries are currently not accessible by observations in the GC. Furthermore, the observed X-ray source overabundance in the central pc \citep{Muno+2005} suggests that compact binaries may reside there as well.

 \citet{Gillessen+2012} has recently reported the discovery of a gas cloud of about 3 $M_{\rm Earth}$, called G2, plunging towards the MBH, with a closest approach distance that brings it as close as $3100$~times the event horizon ($\sim 245$~AU). This discovery generated many models to explain the origin of G2 \citep[e.g.,][]{MCL2012, Burkert+2012, ME2012, Morris+2012, Phifer+2013, Guillochon+2014}. This object gives a rare opportunity to study the dynamics and tidal evolution of a source close to a MBH in real time. Follow-up observations have shown that G2 remained compact in the continuum and that its orbit is still consistent with a Keplerian orbit after periapsis passage \citep{Witzel+2014,Meyer+2014,Valencia+2015}.  Thus, it was suggested that this object is actually a result of a merged stellar binary \citep{Witzel+2014, Sitarski+2016}, a connection previously suggested by \citet{Prodan+2015}. Here we take their study further and couple post main sequence evolution to the dynamical evolution. {The possibility of binary mergers due to perturbations by the MBH was first suggested, in the context of hypervelocity stars, by \citet{Ginsberg+2007}.}

Stellar binaries in the GC form hierarchical triple body systems, with the MBH as the third, distant object. Therefore, stellar binaries are expected to undergo large amplitude eccentricity and inclination oscillations, i.e., the so-called ``eccentric Kozai-Lidov" (EKL) mechanism \citep[see for review:][]{Naoz2016}. Eccentricity excitations in a binary, induced by the MBH, can cause the binary stars to merge {\citep[e.g.,][]{Antonini2010,Antonini+2011}}. However, tidal forces between the binary companions tend to shrink and circularize their orbits, either preventing or severly delaying a merger. Here we take into account post main sequence stellar evolution and show that this, in combination with the EKL mechanism, can cause large fractions of merged systems.  While the mass loss during stellar evolution widens the orbit, the mass loss can also re-trigger the EKL behavior for similar mass binaries, which will lead to large eccentricities \citep[e.g.,][]{Shappee+13, Michaely+14}.  In addition, as the star expands, it may undergo Roche-lobe overflow, especially for binaries with tidally shrunken and circularized orbits, \citep[e.g.,][]{Naoz+2015}. Therefore, contact binaries and merger products can be formed after one of the binary companions has left the main sequence. Merging binaries solves many unanswered questions in the GC context. For example, merging binaries may form rejuvenated products that appear young, which could explain the unexpected population of young stars in the GC \citep{Ghez+2005} and it may change the initial mass function (IMF) of the GC stellar population, which could explain the observed top-heavy IMF in the GC \citep{Lu+2013}.

The paper at first describes the numerical setup of our work (Section \ref{sec:Num}), after which it presents results and an analysis of the binary evolutions observed in our simulations (Section \ref{sec:res}). This is followed by a discussion of our results in Section \ref{sec:dis}. Appendix \ref{App:AppendixA} is attached to discuss the evaporation timescale.


\begin{figure}
\hspace{-0.075\columnwidth}
\includegraphics[width=1.1\columnwidth ]{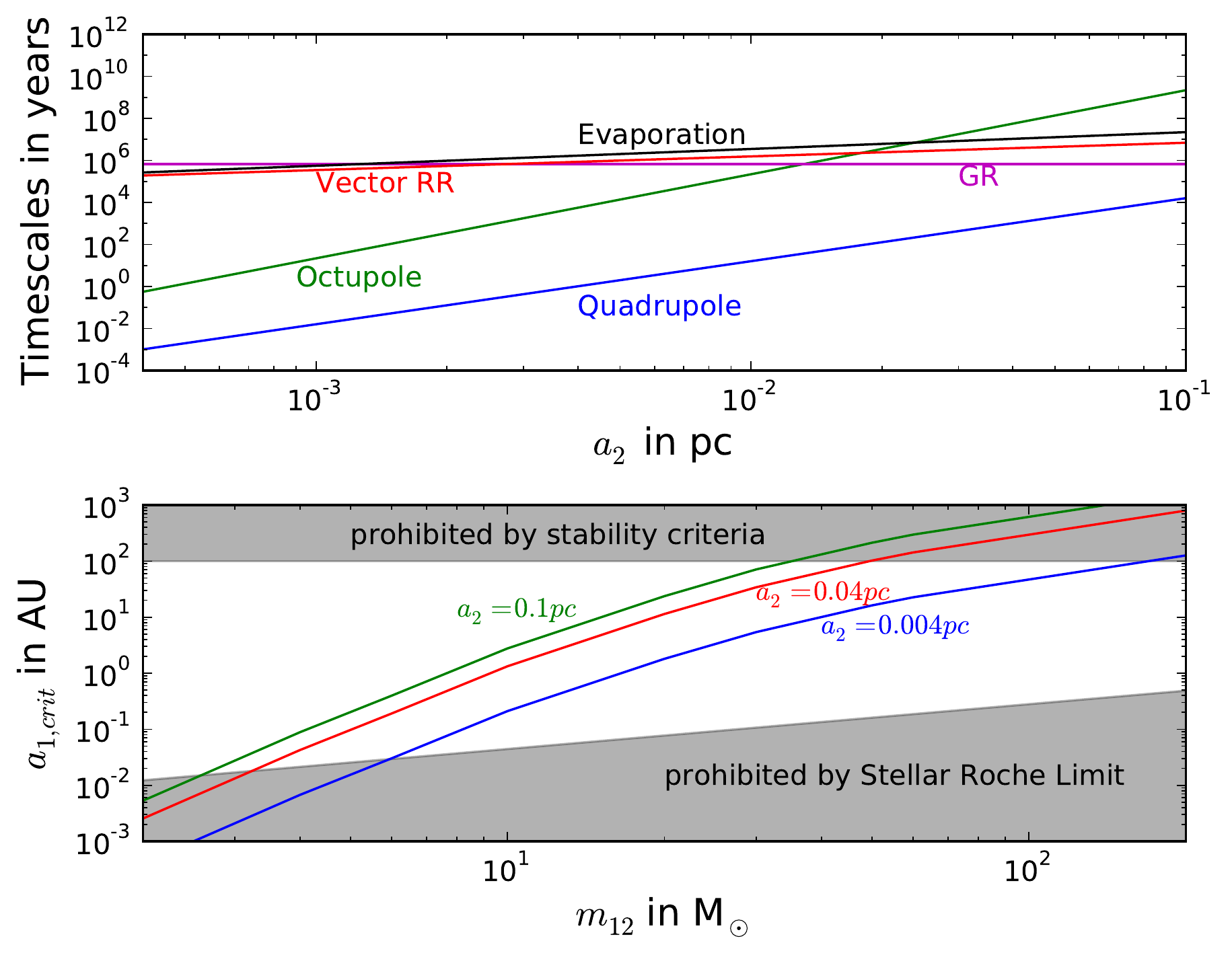}
\caption{{\bf Timescales of different physical effects and critical $a_1$ values for stellar evolution.} Plotted in the top panel are the periapsis precession timescales associated with the EKL mechanism (quadrupole in blue, octupole in green), GR in the inner binary (magenta), vector resonant relaxation (VRR, red), and the evaporation timescale for an example binary of $2$ $M_\odot$ total mass and $a_1 = 3$~AU as a function of $a_2$. Not shown is the outer orbit precession timescale induced by general relativity by the MBH of $\sim 10^5$~yrs, which is not of primary concern for our calculations. The scalar resonant relaxation timescale is in general between $10^{9}$ and $10^{11}$~yrs and can dip down to $~10^7$~yrs around $a_2=0.007$~AU due to the cancellation of the relativistic precession by Newtonian precession \citep[see][]{KocTrem2011}. We omit it here to avoid clutter. EKL oscillations can only occur if their associated timescales are shorter than the other precession effects, since they counteract the buildup of large eccentricities. Note that the evaporation timescale is comparable with the VRR timescale, which allows us to ignore VRR for our calculations\protect\footnotemark. Plotted in the bottom panel are the critical maximum binary separations, $a_1$, allowed for stellar evolution to take place before the binaries evaporate. The gray areas denote $a_1$ values prohibited by the stellar Roche limits (bottom), or by the stability criteria of our initial conditions (top).}
   \label{fig:timescales}
\end{figure}

\footnotetext{Note that the precession of the outer orbit due to general relativity \citep{Naoz+12GR} or due to the precession from the spherical star cluster (i.e, Newtonian precession, \citealt{KocTrem2011}) may have quantitative effects on a single system but will not change the overall statistics. We tested the effect of GR effects on the outer orbit and did not find any differences in our results.}

\section{Numerical Setup}\label{sec:Num}

 The numerical setup of the systems is chosen to be consistent with binaries in the field. The mass distribution of the primary stellar binary member is taken as a Salpeter function with $\alpha=2.35$ with a minimum mass of $1$ and a maximum of $150$ M{$_\odot$} (we denote this mass as $m_1$), while the mass ratio to the secondary mass ($m_2$) is taken from \citet{Duquennoy+91}. This choice of numerical binary setup differs from  \citet{AntPer2012} and \citet{Prodan+2015} as we do not have a large fraction of $m_1=m_2$ systems. {While some studies have suggested that stellar twin binaries are relatively common \citep[see, for example][]{PinsonneaultStanek2006, KobulnickyFryer2007}, more recent works suggest that stellar twins do not form a relevant part of the binary population \citep{Lucy2006,Sana+2012}.} Therefore, avoiding exact stellar twins is a more realistic choice for the mass distribution. Additionaly, since the octupole level of approximation goes to zero for systems that have an exact symmetry \citep[e.g.,][]{Naoz2016}, relaxing the symmetry allows the systems to undergo large eccentricity excitations even for similar masses once the stars begin to evolve and lose mass \citep{Shappee+13,Michaely+14}.

\begin{figure}
\hspace{-0.075\columnwidth}
\includegraphics[width=1.1\columnwidth ]{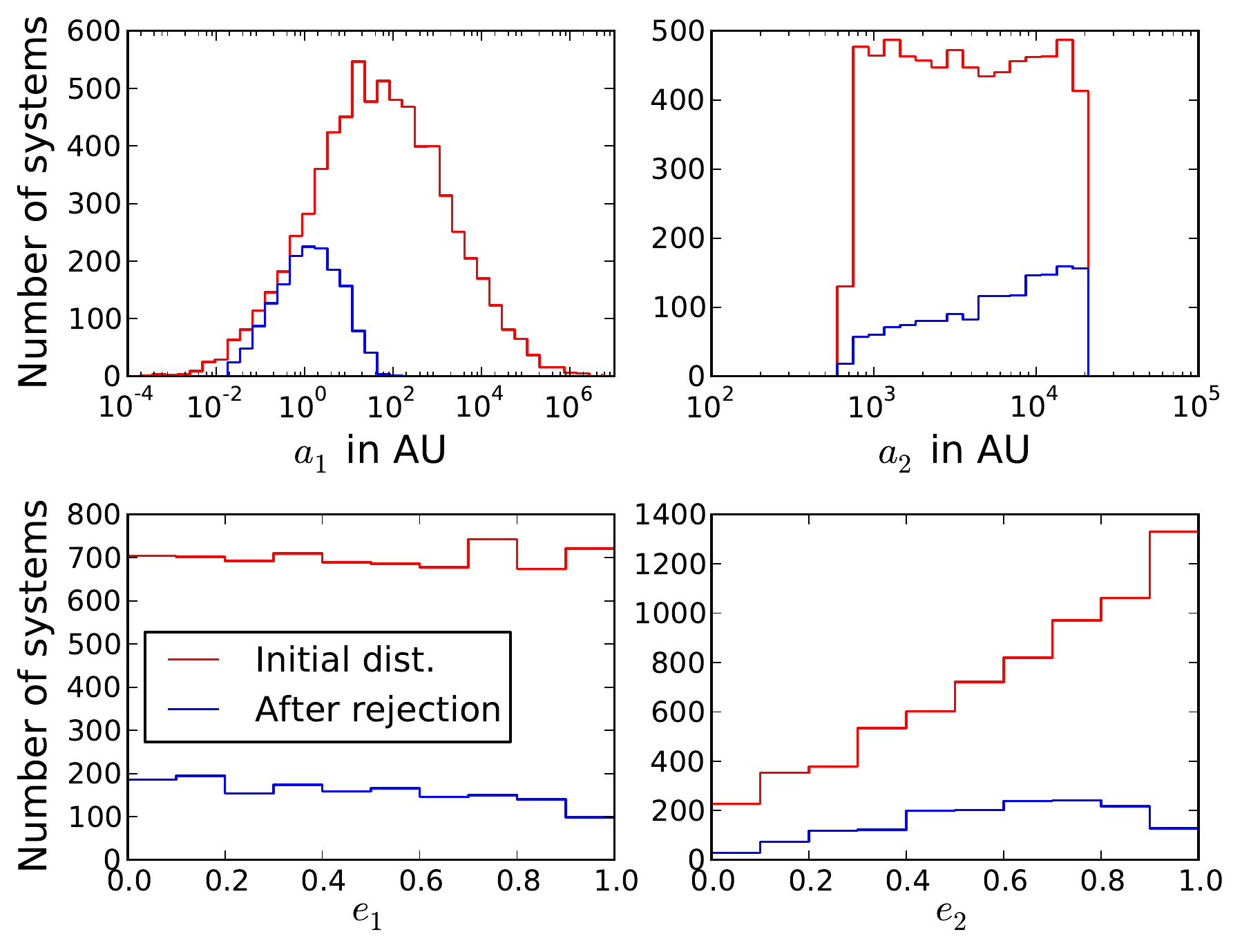}
\caption{{\bf Parameter distributions before and after stability test.} Shown are the distributions of $a_1$, $a_2$, $e_1$, and $e_2$ before and after the stability test. Note that the $a_1$ distribution remains log-normal after the stability test, its peak is simply shifted to a smaller value. The new $a_2$ distribution has a small preference for larger $a_2$ values compared to the flat original distribution. The $e_1$ distribution remains flat, while the $e_2$ distribution preferentially suppresses large $e_2$ values of the originally thermal distribution. The inclination distribution (not shown) did not change at all. 
}
   \label{fig:distributions}
\end{figure}

The mass of the MBH, {$m_{\rm{MBH}}$}, remains fixed at $4\times10^6$~M{$_\odot$} \citep[following][]{Ghez+2005,Gillessen+2009} for all systems since we are mostly interested in the results for Sagittarius A{$^\star$}. The eccentricity distribution is taken as uniform from $0$ to $1$ for the stellar binary (inner) orbit ($e_1$), following \citet{Raghavan+10}, while it is taken as thermal for the MBH (outer) orbit ($e_2$) \citep{Jeans1919}. The mutual inclination between the inner and outer orbit is assumed to be uniformly distributed in cosine. The argument of periapsis is taken from a uniform distribution for both orbits, as is the angle of the stellar spin axis. The outer orbital period is distributed uniformly in log space, with an assumed minimum of $\sim 9$~yr corresponding to a semi-major axis ($a_2$) of $\sim 700$~AU (the closest known star to Sagittarius A{$^\star$}, SO-2, has an orbital period of $\sim 15$~yrs, \citealt{Ghez+2005}) and an assumed maximum of $\sim 1500$~yrs or $0.1$~pc, due to the conflicting timescales for orbits larger than that, as seen in Figure \ref{fig:timescales}. The inner orbital semi-major axis ($a_1$) and period is taken from a log-normal distribution with a peak at $\sim 170$~yr and the one sigma interval going from $\sim 0.9$ to $\sim 35000$~yr \citep{Duquennoy+91}.

We require these randomly generated systems to satisfy dynamical stability so that we can separate the long-term secular effects.  Similarly to \citet{Naoz+14stars}, the first condition requires that the two stars are initially outside their binary compaion star's Roche limit, otherwise  the inner stars suffer a merger before the tertiary can act.  The second condition requires that the system is hierarchical enough to allow usage of the  EKL equations. This means that the systems have to fulfill the following criterion:
\begin{equation}\label{eq:epsilon}
\epsilon=\frac{a_1}{a_2 }\frac{e_2 }{1-e_2^2}<0.1 \ ,
\end{equation}
where $\epsilon$ is a measure of the relative strengths of the octupole and quadrupole effects on the orbital dynamics \citep[e.g.,][]{Naoz2016}\footnote{Numerically, this is very similar to the \citet{Mardling+01} nominal triple stars stability criterion, as shown in  \citet{Naoz+12GR}.}. Similar to the first condition, which requires that the inner binary members are not already crossing their partner stars' Roche limit, the third criterion requires that the inner binary does not cross the MBH's Roche limit, in the form of:
\begin{equation}\label{eq:BHroche}
\frac{a_2}{a_1}>\left( \frac{3m_{\rm{MBH}}}{m_{\rm{binary}}} \right)^{1/3}\frac{1+e_1}{1-e_2}
\end{equation}
Even though we reject systems that violate Equation \ref{eq:BHroche} from the set of initial conditions, some systems cross the MBH's Roche limit later during their evolution. We keep track of those and will analyze them separately.

After applying these stability tests many of the original systems were rejected. Out of  $7000$ systems with the parameters mentioned above,  we are left with $1570$  systems. Note that the inner  period initial conditions distribution following \citet{Duquennoy+91} is probably unrealistic for the condition for star formation in the GC. Specifically, it is unlikely that star formation episodes at the GC will lead to the formation of binaries with separations larger than $100$~AU, as suggested by  \citet{Duquennoy+91}. Furthermore, the limited parameter space available for forming stable binary systems should also severely limit the possibility of forming triple or higher multiplicity stars in the GC, which we therefore ignore in our model. 

The exact distribution for the inner binary period is unknown and thus we follow this recipe for initial conditions generation. We ensure the stability of the systems by our stability criterions. We thus caution the reader from interpreting the rejection method of our initial conditions as a physical process.
The orbital parameter distributions before and after our rejection process are depicted in Figure \ref{fig:distributions}.

We solve the hierarchical triple secular equations up to the octupole level of approximation \citep[as described in][]{Naoz+11sec, Naoz2016}, including  general relativity (GR) effects for both the inner and outer orbit \citep{Naoz+12GR}\footnote{{{ We note that the GR precession of the  outer orbit has an insignificant effect on the dynamical evolution and can be neglected. }} } and static tides for both members of the stellar binary \citep[following][]{Hut,1998EKH}. See \citet{Naoz2016} for the complete set of equations. We also include the effects of  stellar evolution on stellar radii, masses, and spins following the stellar evolution code {\tt SSE} by \citet{Hurley+00}. {{ The octupole level code with post main sequence effects was successfully tested in a previous  study \citep{Naoz+2015}. Our code, which couples the secular evolution with the  physical processes mentioned above, allows us to go beyond the initial study of the dynamical evolution of binaries at the GC by \citet{AntPer2012} and \citet{Prodan+2015}.  Specifically, the post main sequence evolution adds significant contributions to merging the binaries. } }

We have two sets of Monte-Carlo simulations {{adopting}} two different tidal efficiencies.  One series of systems has efficient tides, which is achieved by using a constant viscous time of $1.5$~years ($750$~systems); the other set of systems has less efficient tides with a constant viscous time of $150$~years ($1570$ systems). The different assumptions for the viscous time cause variations in the formation likelihood of tight, circular binaries. The less efficient tides are probably  more realistic \citep[e.g.,][]{Hansen10} and we use a larger sample size for those runs in order to obtain better statistics. {Unless explicitly stated otherwise, all results discussed in this work refer to the more realistic less efficient tide scenario.}

Eccentricity excitations in the inner binary orbit will take place if the shortest timescale associated with the EKL mechanism, i.e., the quadrupole timescale, is shorter than the  precession of the periapsis due to short range forces.  Such short range forces come from GR and the precession of the nodes due to oblate objects from static tides or rotating objects \citep[e.g.,][]{Naoz2016}. The octupole  level of approximation assisted timescale gives a sense of the timescale required to pump the eccentricity up to extreme eccentricity spikes. In Figure \ref{fig:timescales} we show the quadrupole, octupole and GR timescales for a nominal example system \citep[see][ for the relevant timescales]{Naoz2016}. This shows that the quadrupole timescale is shorter than the GR timescale for the considered values for $a_2$, thus allowing eccentricity and inclination oscillations to occur.

Binary mergers in dense stellar groups can be prevented through close encounters with other stars, which can change the orbital parameters of the binaries. The change depends to a large degree on whether the binary's orbital energy is larger or smaller than its center-of-mass motion energy compared to other cluster stars, which are known as {\bf hard} or {\bf soft} binaries, respectively. Binaries in the GC are generally soft and are expected to become less bound through stellar encounters, until they disassociate or ``evaporate".  A derivation for the evaporation timescale can be found in \citet{Binney+Tremaine1987}. For completeness, the derivation of the evaporation timescale with the total mass of the system is presented in  Appendix \ref{App:AppendixA}.  The evaporation timescale for a nominal example system is also shown in 
Figure \ref{fig:timescales}. 
However, some binaries in the GC can be hard, especially if their orbital separation is reduced by tidal effects. Hard binaries can actually form even tighter, harder binaries through stellar interactions and their evaporation timescale becomes exponentially longer with their hardness \citep[see, for example,][]{Heggie1975, Hut1983, Hut+1983, Heggie+1996}. GC binaries need to be very tight in order to be hard, $a_1$ needs to be on the order of $\sim 0.1$~AU or less. At such separations the tides of the binary companions become very strong. We therefore assume that such tight binaries survive evaporation effects for at least $10$~Gyrs, see  Appendix \ref{App:AppendixA}. 

Mergers through EKL oscillations can only occur if the timescale of quadrupole effects is shorter than the evaporation timescale. However, post main sequence stellar evolution can also lead to mergers, which is likewise limited by the survival time of the binary. The bottom panel of Figure \ref{fig:timescales} shows the maximum binary separation possible in order for post main sequence stellar evolution to occur before the binary evaporates.

Another important physical process that takes place in the GC is vector resonant relaxation (VRR). This process changes the direction of the outer orbit's angular momentum, but not its magnitude. Since efficient eccentricity excitations due to the EKL mechanism requires large mutual inclinations between the inner and outer orbits\footnote{If both the inner and the outer orbits are initially  eccentric, further eccentricity excitations can take place in a nearly coplanar configurations \citep{Li+13}.  }, VRR will change the inclination and will effectively  refill the available phase space  that  allows large eccentricity oscillations.  Binaries that did not undergo substantial tidal effects nor merged, can, due to the VRR process, change their inclination relative to the MBH such that they leave the favorable EKL regime. The associated timescale is shown in Figure \ref{fig:timescales}  \citep[e.g.,][]{Hopman+2006,KocTrem2011,KocTrem2015}.  As depicted in this example, the VRR timescale is comparable to the evaporation timescale for most parts of the parameter space, but becomes shorter than the evaporation timescale beyond $\sim 0.1$~pc. Thus, we neglect the VRR effects in our calculations, but limit them to the inner 0.1 pc of the GC. However, we note here that the VRR timescale is in general too long to have an effect on EKL induced mergers; see Section \ref{sec:res} for an explanation in light of the results of our calculations.

The integration time for each binary is determined by its evaporation timescale (see Appendix \ref{App:AppendixA}). If a binary became tightly locked, and thus decoupled from the gravitational perturbations of the MBH, we recalculate the evaporation time (we set the evaporation time to $10$~Gyrs for hard binaries) and continue the post main sequence evolution using the binary stellar evolution code {\tt BSE} \citep{Hurley+02} until the binary either merged or evaporated. Our maximum evolution time is $10$~Gyrs.

\section{Results}\label{sec:res}

 {{ During the EKL evolution, the inner orbit eccentricity is excited to extremely high values \citep[e.g.,][]{Tey+13,Li+13,Li+14} that can result in crossing of the Roche limit, which may drive the two stars into merging \citep[e.g.,][]{Naoz+14stars}.
 The Roche limit is  defined by
\begin{equation}\label{eq:aRoche}
a_{{\rm Roche},ij}=\frac{R_j}{\mu_{{\rm Roche},ji}} \ ,
\end{equation}
where $R_j$ is the radius of the star at mass $m_j$ and, following \citet{Eggleton83}, $\mu_{{\rm Roche},ji}$ is a  dimensionless    number
  \begin{equation}
\mu_{{\rm Roche},ji}=0.49\frac{ ({m_j} /{m_i})^{2/3} }{0.6 (m_j/m_i)^{2/3}+\ln (1+(m_j/m_i)^{1/3})} \ .
\end{equation}
If the inner binary pericenter distance, $a_1(1-e_1)$, becomes smaller than either $a_{{\rm Roche},12}$ or $a_{{\rm Roche},21}$, we assume that one of the stars will overflow its Roche lobe. We stop the calculation here and identify this system as a merger candidate. 


\begin{figure*}
\hspace{0.0\linewidth}
\includegraphics[width=0.9\linewidth ]{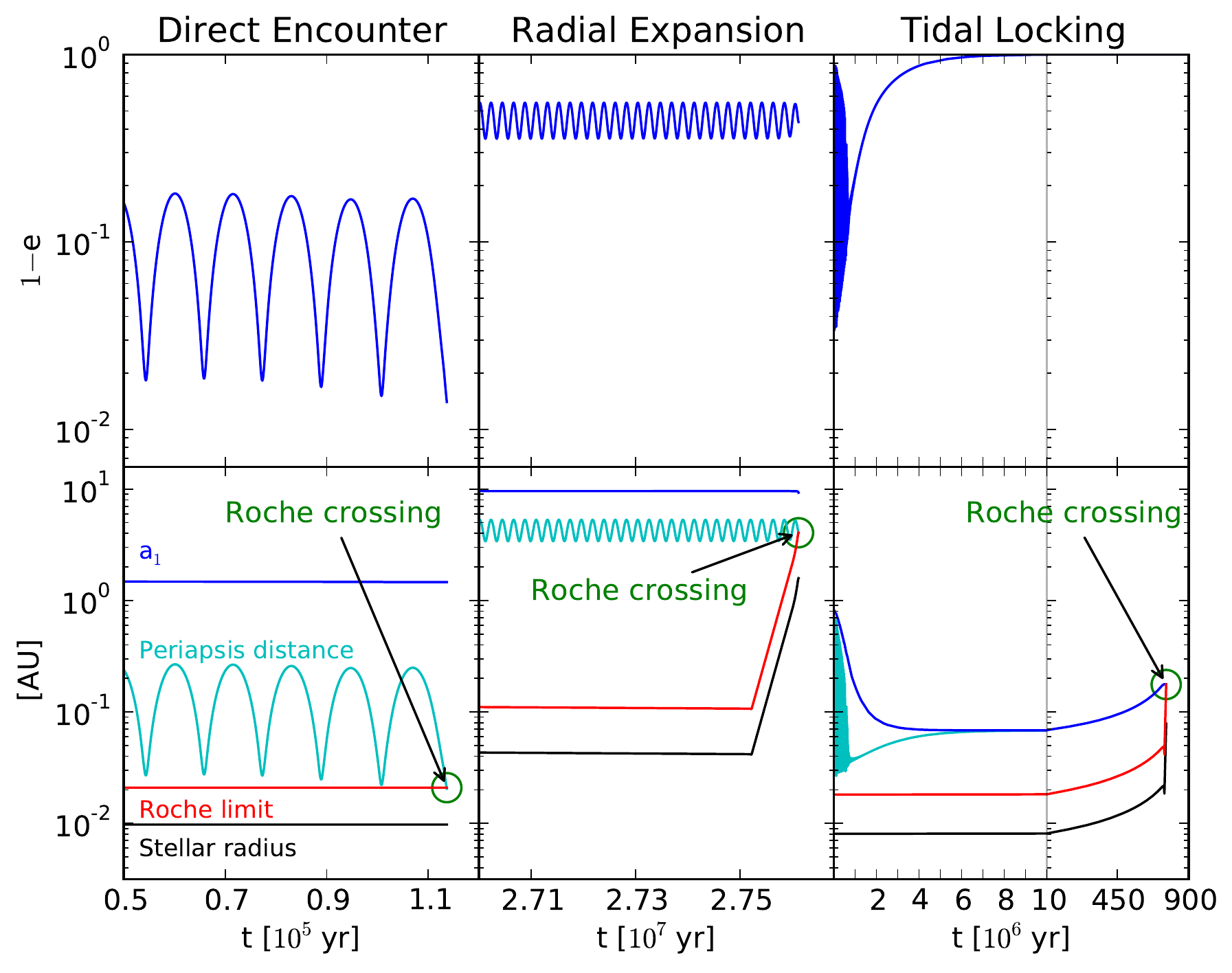}
\caption{{\bf Example Evolution for the three merger scenarios} The plot shows the evolution of eccentricity, semi-major axis, periapsis distance, Roche limit, and stellar radius for three example systems that lead to mergers. The left column shows a {\bf direct merger} caused by high orbital eccentricity due to EKL cycles, the center column shows a {\bf radial merger} caused by the rapid expansion of the larger stellar companion in the binary due to stellar evolution, and the right column shows a {\bf tidal merger}, a system that becomes tidally locked before it can evaporate and will survive long enough to merge due to radial expansion similar to the center column's example system. Note that the systems in the center and the right column are examples of regular contact binary mergers, while the left column's system could be an example of a direct collision or otherwise high-velocity merger with a much more complex binary evolution, which is beyond the scope of this work. After $6$~Myrs of evolution we register mostly direct mergers and a few radial mergers, after $10$~Gyrs $35 \%$~of all mergers are direct mergers, $10 \%$~are radial mergers and $55 \%$~are tidal mergers. }
   \label{fig:largeplot}
\end{figure*}

 
 We categorize the outcomes of the binaries' evolution into three main cases:
  \begin{enumerate}
   \item {\it Tidally Locked Binaries.} The tidal interactions of the binaries will in some cases lead to circularization and shrinking of binary orbits. The stars' spin periods and orbital period will in this case synchronize, tidally locking the binary. These tight, tidally locked binaries are an important class of objects, since the tightening of their orbits can significantly delay evaporation of the binaries. Since strong tidal evolution significantly slows down the computation, we stop the simulation if the orbit is sufficiently tight ($a_1<0.1$ AU), circularized ($e_1<10^{-5}$), and if the spin and orbital periods have synchronized to approximately reach the expected spin rate $\Omega$ {(within a fractional difference of $10^{-5}$)}   \citep[e.g.,][]{Fabrycky+07}:
  \begin{equation}\label{eq:Fabr_spin}
\Omega=2\frac{2\pi/P_{\rm{in}}}{\cos\psi+\sec\psi}
\end{equation}
Here, $P_{\rm{in}}$ is the inner binary period and $\psi$ is the orbital obliquity.

Tidally locked binaries are a transient phase with their duration determined by the main sequence lifetime of the more massive star. If the post main sequence radial expansion occurs before the binary is somehow disrupted, the stars will merge (see below). We found that after $6$~Myrs of evolution about $10 \%$ of the systems became tidally locked (using the low-efficiency tides model; see the right hand column of Figure \ref{fig:largeplot} as an example). The system undergoes large eccentricity excitations while tidal interactions are slowly shrinking and circularizing the orbit. As discussed below, after post main sequence evolution many of these tidally locked systems become merged binaries.

  \item {\it Merged Binaries and G2-like objects}.
 There are three channels for merging binaries. The first channel consists of Roche limit crossing at the periapsis via eccentricity excitations due to EKL dynamical evolution (depicted in the left column of Figure \ref{fig:largeplot}), which we coin as {\bf ``direct mergers"}\footnote{ This channel was noted already by \citet{Antonini2010}, \citet{AntPer2012} and \citet{Prodan+2015}, although their abundance of occurrences  was smaller due to a larger abundance of twin stars.}.
  The second channel consists of {\bf ``radial mergers"} in wide binaries due to Roche lobe overflow during the post main sequence expansion of the more massive binary companions (see middle panels in Figure \ref{fig:largeplot}).  The final channel takes place in the post main sequence evolution of a tidally locked binary. This type of system has decoupled  from the MBH gravitational perturbations and thus the subsequent evolution of the binary can be followed using BSE. We recalculate the evaporation time for this system and allow for stellar evolution to take place. As the more massive star evolves beyond the main sequence, its radius begins to expand and overflow its Roche lobe (see left panels in Figure \ref{fig:largeplot}). We find, as expected, that most (about $95 \%$) of the tidally locked binaries end up as merged systems after $10$~Gyrs of evolution, while the remaining unmerged systems have just started expanding in radius as they are close to $1$~$M_\odot$ in mass. We coin this merging channel as {\bf ``tidal mergers"}.

Considering that the last  star formation episode  in the GC happened about $6$~Myrs ago, the merged binaries would still be in a morphed extended phase, which may be observable as G2-like objects \citep{Witzel+2014, Sitarski+2016}. We find that at this timescale nearly all merged systems would be direct mergers, with only a few radial mergers just occuring from the most massive stars. No tidal mergers would have occured yet, as can be seen in Figure \ref{fig:timeVSmass}. {As mentioned before, binaries are registered as mergers if they crossed their Roche limit, and even direct mergers probably represent at most grazing encounters and not head-on collisions. The physical merging into a single star will occur after an extended merging process that may last for a few Myrs (for circular orbits, as found using BSE). During the merging process, the two stars enter a dusty red phase \citep[e.g.,][]{Tylenda+2011_1,Tylenda+2011_2,Tylenda+2013, Nicholls+2013}, which may be identified as an IR excess source \citep[e.g.,][]{Witzel+2014}. Once the two stars have merged into one, the product  will readjust to form a stable star on the order of a Kelvin-Helmholtz timescale. This entire process gives enough time to observe a system that began merging within the last $6$ Myrs as a G2-like object today.}

{We postulate that all mergers will, in the short term, appear as G2-like IR excess sources, but after they have settled down the newly formed stars will look rejuvenated, younger than the other stars in their population, due to their now enlarged mass and previous slower main sequence evolution speed. If a star formation burst indeed took place about $6$~Myrs ago, then we expect that the G2-like objects are direct mergers still going through the merging process. However, our results also suggest a possible connection between the young stellar population and the merger products. The young stellar population in the GC consists of fairly massive stars and could be a sample of settled down mergers \citep[e.g.,][]{Ghez+2003}. Specifically, the S-stars may be consistent with the late mergers in our simulations, which would indicate that a star formation period took place significantly earlier than $6$~Myrs ago, as these binary mergers would have needed time to settle down to their current state. Of course, it is possible that several star formation episodes have occurred, which would then allow G2-like objects to be formed from recent episodes, while older episode mergers would appear rejuvenated today.} We find that {after $10$~Gyrs of evolution} the direct mergers {constitute} about $35 \%$ of total merged population, while the radial mergers make up about $10 \%$ and the tidal mergers about $55 \%$.

As an overview, merging binaries will either undergo direct mergers or mergers due to stellar evolution (either radial mergers or tidal mergers). This is depicted in Figure \ref{fig:timeVSmass}, where the stellar evolution induced mergers are mostly located close to the end of the stellar mass lifetime.
 
\begin{figure}
\hspace{-0.075\columnwidth}
\includegraphics[width=1.1\columnwidth ]{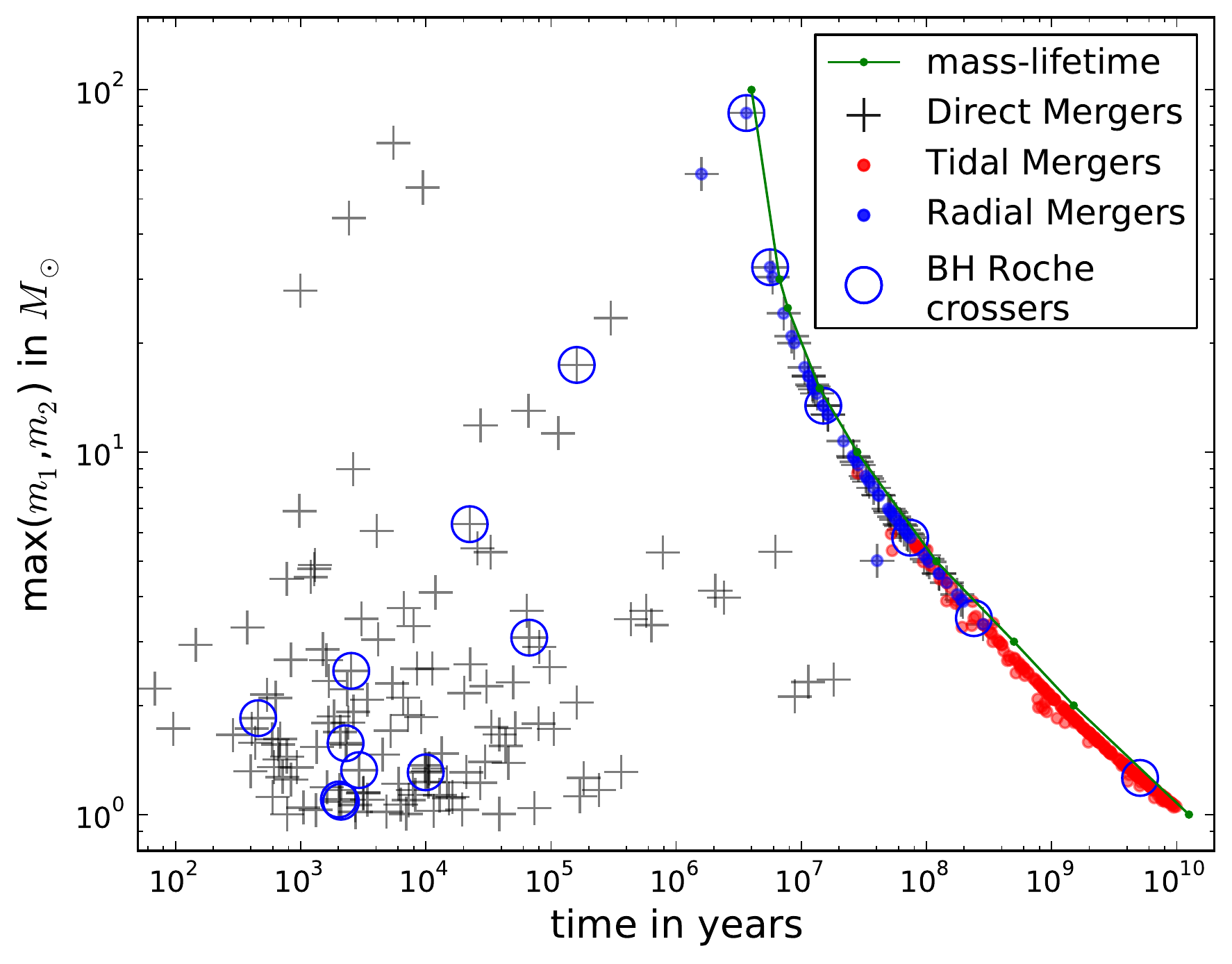}
\caption{{\bf Time of mergers compared to main-sequence lifetimes.} Plotted are the times that any particular merger happened vs. the masses of the larger companion in the binary. The main-sequence lifetime for the range of $1$ to $100$~$M_\odot$ is plotted in green, which allows us to compare if a merger happened mostly due to EKL induced large eccentricities or rather due to radial expansion of the more massive companion at the end of its main sequence lifetime. Black crosses stand for mergers that were directly registered by our simulation, while red dots stand for those registered tidally locked systems that we expect to merge once inflation begins, based on the binaries' separation at tidal locking and expected radial size during inflation. Blue dots
represent mergers of not tidally locked systems that were caused by the radial expansion of one of the binary companions due to stellar evolution. Blue circles indicate binaries registered as mergers that have crossed the MBH Roche limit during their evolution.
 }
   \label{fig:timeVSmass}
\end{figure}

\begin{figure*}
\begin{center}
\includegraphics[width=0.7\linewidth ]{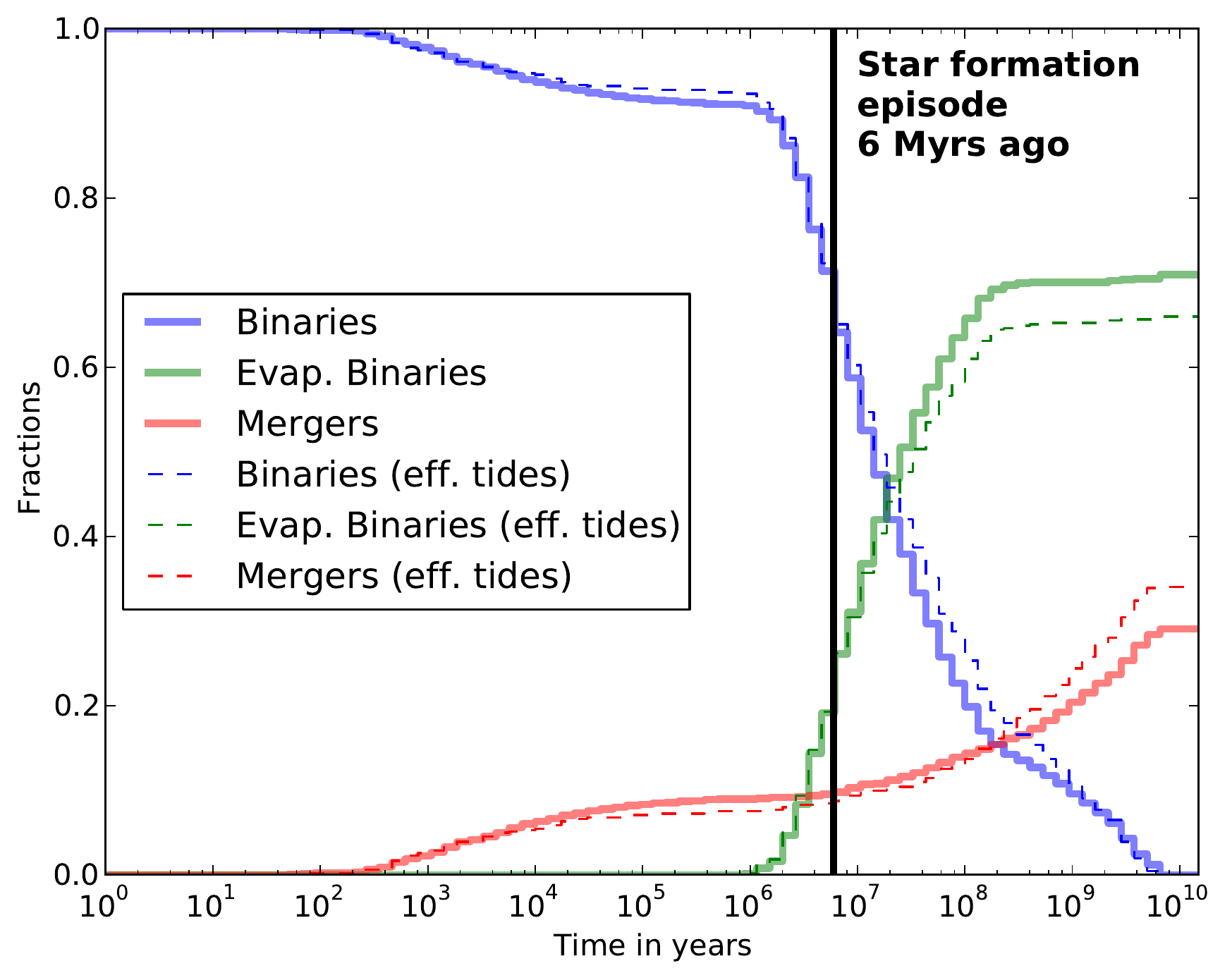}
\caption{{\bf Binary, evaporated, and merger fractions over time.} Plotted are the fractions of binaries, evaporated binaries, and merged binaries as a function of time. The solid, colored lines show the results for the (normal) inefficient tide model, while the dashed lines show the results for the efficient tide model. Note that the only significant difference between those two models is in the long term expected total fraction of mergers due to the larger probability for the formation of tidally locked systems. In the short to medium term the results are virtually identical. The black vertical line marks the age of the young stellar population in the GC, determined to be $6$~Myrs by \citep{Lu+2013}. Note that the binaries and evaporated binaries distributions of the two models begin to diverge only for stellar populations older than that, with fewer evaporated binaries in the efficient tides case. Thus, the tidal model is not of dominant importance in order to predict the fractions of mergers and binaries for the young stellar population. The different tidal strengths are mostly just influencing the formation of tidally locked systems, which can become merger products at later times.}  
   \label{fig:fracVStime}
   \end{center}
\end{figure*}

  \item {\it Evaporated binaries}. Binary systems that never merged or tidally locked are assumed to evaporate after their evaporation time is reached and the simulation is halted. Stars that tidally locked, but did not evolve and merge before the end of their newly calculated evaporation time, are likewise assumed to evaporate. Furthermore, systems that, at any point in their evolution, crossed the MBH Roche limit due to changes of the inner binary eccentricity $e_1$ (see Equation \ref{eq:BHroche}) are assumed to evaporate as well. Binaries that cross the MBH Roche limit will generally become unbound, while those in the double loss cone have a high chance of merging, as shown by \citet{MandelLevin2015} (the double loss cone is defined as the set of orbits around the MBH for which a binary is supposed to become unbound, while the individual binary companions are also supposed to be tidally disrupted). We have $104$ of all {$1570$ calculated} systems, among them $18$ merging systems, that cross the single loss cone during their evolution. We therefore assume that those $18$ systems evaporate instead of merge. This reduces the total number of merged system only insignificantly, on the order of $4 \%$ of total mergers, or $\sim 1 \%$ of all systems. None of our systems cross the double loss cone, since our intial conditions already rejected all systems that would bring the binary so close to the MBH. The systems also cannot evolve to cross the double loss cone later, since $a_2$ does not change. 
  
\end{enumerate}
 
 As depicted in  Figure \ref{fig:fracVStime}, the fraction of binary stars declines over time both due to merging and evaporating. \citet{Lu+2013} found that  the last star formation episode in the inner parts of the GC  took place a few million years ago (adopted here as $6$~Myrs ago). Considering this timescale, about $70\%$ of the initial binary fraction is still expected to be present at the GC.  Furthermore, we expect that about $13\%$ of the initial binary population has merged, and thus may be detectable as G2-like objects. {\it Therefore, we refer below to all merged products after $6$~Myrs as G2-like candidates. }

Interestingly, the  different tidal assumptions are indistinguishable at  the $6$~Myrs mark. Prior to that the {\bf direct mergers} channel is the only operating channel. The efficiency of this channel is then reduced (see also Figure \ref{fig:timeVSmass}), and the merger rate stalls. This leads to an apparent plateau in the fraction of mergers at this timescale. A few tens of Myrs later, the massive stars start to leave the main sequence and inflate in size. Thus, the merger rate increases again by allowing {\bf radial mergers} to occur. As the tidal evolution equations are highly sensitive to the stellar radius \citep[e.g.,][]{Naoz2016}, the more efficient tides tend to produce slightly more merged systems over long times by forming more tidally locked and circularized systems and enabling the {\bf tidal mergers} channel.

The binary population of the earliest star formation episode roughly $10$~Gyrs ago has been completely depleted and the binaries have either evaporated or have merged as depicted in Figure \ref{fig:fracVStime}. The binary merger rate at late times is dominated by the main sequence lifetime of the more massive binary companion in the tidally locked binaries. Since these binaries are hard we assumed a survival time of $10$~Gyrs, consistent with the estimated ionization rate (see Appendix \ref{App:AppendixA}). After this time basically all of the hard binaries have merged and all remaining soft binaries have evaporated, which leads to the second plateau in the merger rate. About $29\%$ of the initial binary population has merged after  $10$~Gyrs.  Thus, we expect that the old stars will harbor little to no binary systems. However, we expect that a significant  fraction of this population will be the result of  merged systems and may seem younger in comparison. A similar argument was raised for the blue stragglers population \citep{PF09,Naoz+14stars}. }}

\begin{figure}
\hspace{-0.075\columnwidth}
\includegraphics[width=1.1\columnwidth ]{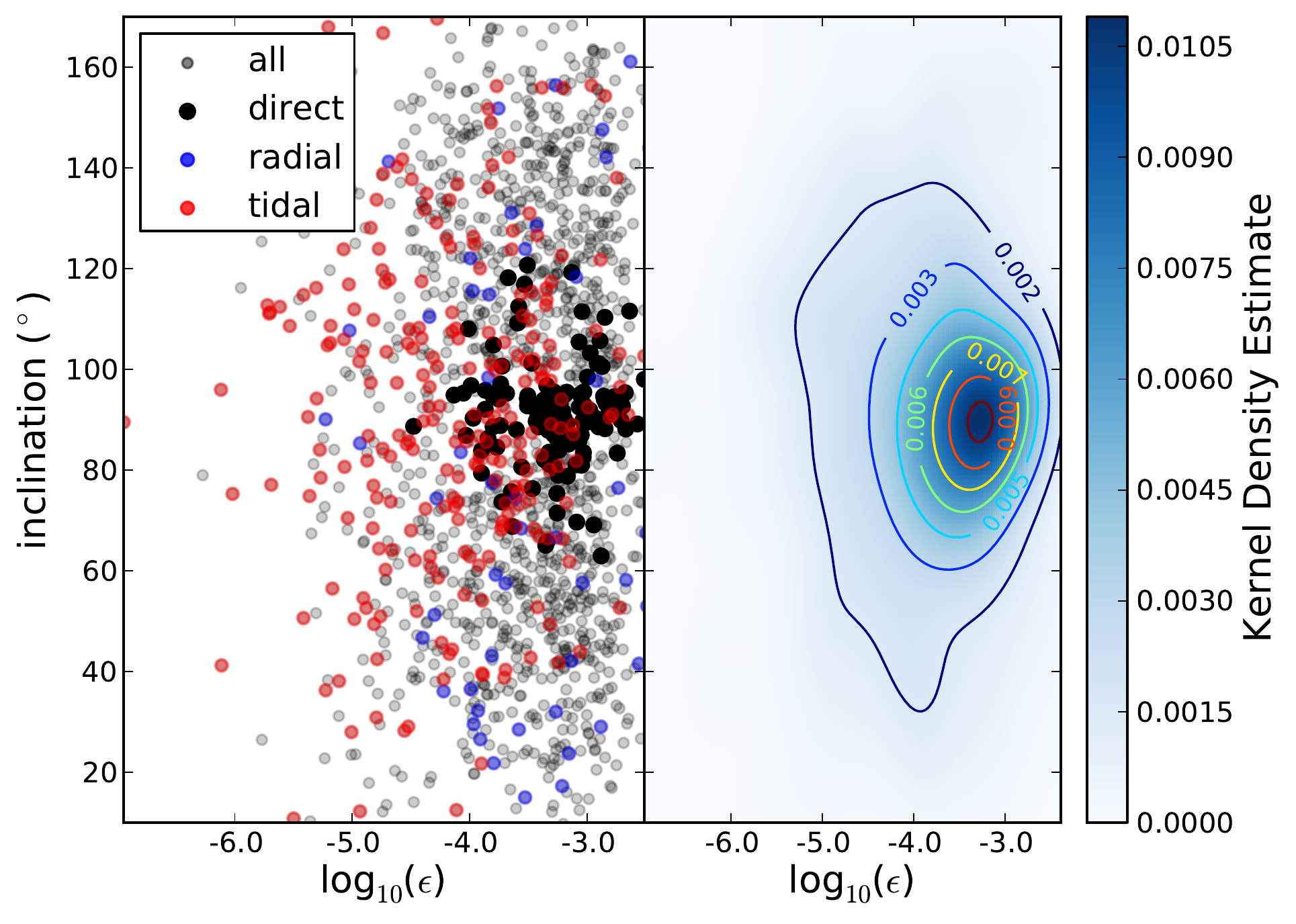}
\caption{{\bf Merger distribution and Gaussian kernel density estimate (KDE) of the mergers as a function of inclination and epsilon.} Plotted in the left panel is the distribution of all systems (in gray) vs. the distribution of mergers (direct in black, radial in blue, tidal in red), and plotted in the right panel is the gaussian KDE, or the smoothed density of all mergers in inclination-epsilon space, assuming an intrinsic gaussian distribution. The colorbar shows the density, while the contours enclose certain density levels (see labels). While we do not expect the intrinsic distribution of mergers to be truly gaussian, the KDE helps to highlight the strong concentration of mergers towards $90^\circ$ inclination and high epsilon values due to direct mergers (for the definition of epsilon, see Equation \ref{eq:epsilon}.}
   \label{fig:incVSeps}
   \vspace{2cm}
\end{figure}

 As expected from the EKL mechanism, merged binaries are   preferentially found in systems with large mutual inclination (however, a large range is allowed), and with a stronger octupole contribution (estimated as $\epsilon$), as  seen in Figure  \ref{fig:incVSeps}. The left panel shows the distribution of all systems and the merged systems in inclination-epsilon space, while the right panel shows the Gaussian kernel density estimate (KDE) of the merger distribution. The KDE can be understood as a smoothed 2D density distribution (a smoothed histogram of mergers per area in the plot's space), assuming an underlying Gaussian distribution. It highlights the strong concentration of mergers at high epsilon values and towards $90^\circ$ inclination.

We note the binary inclination with respect to the MBH is sensitive to the VRR timescale, which will alter the outer orbit angular momentum orientation. This cannot really decrease our merger rate as the direct mergers take place on much shorter timescales than the VRR effects. Furthermore, tidally locked systems are decoupled from the tertiary and are thus insensitive to the outer orbit orientation. Finally, the radial mergers may be marginally effected, however, VRR will refill the EKL high inclination parameter space and can thus retrigger eccentricity excitations. 

\begin{figure}
\hspace{-0.075\columnwidth}
\includegraphics[width=1.1\columnwidth ]{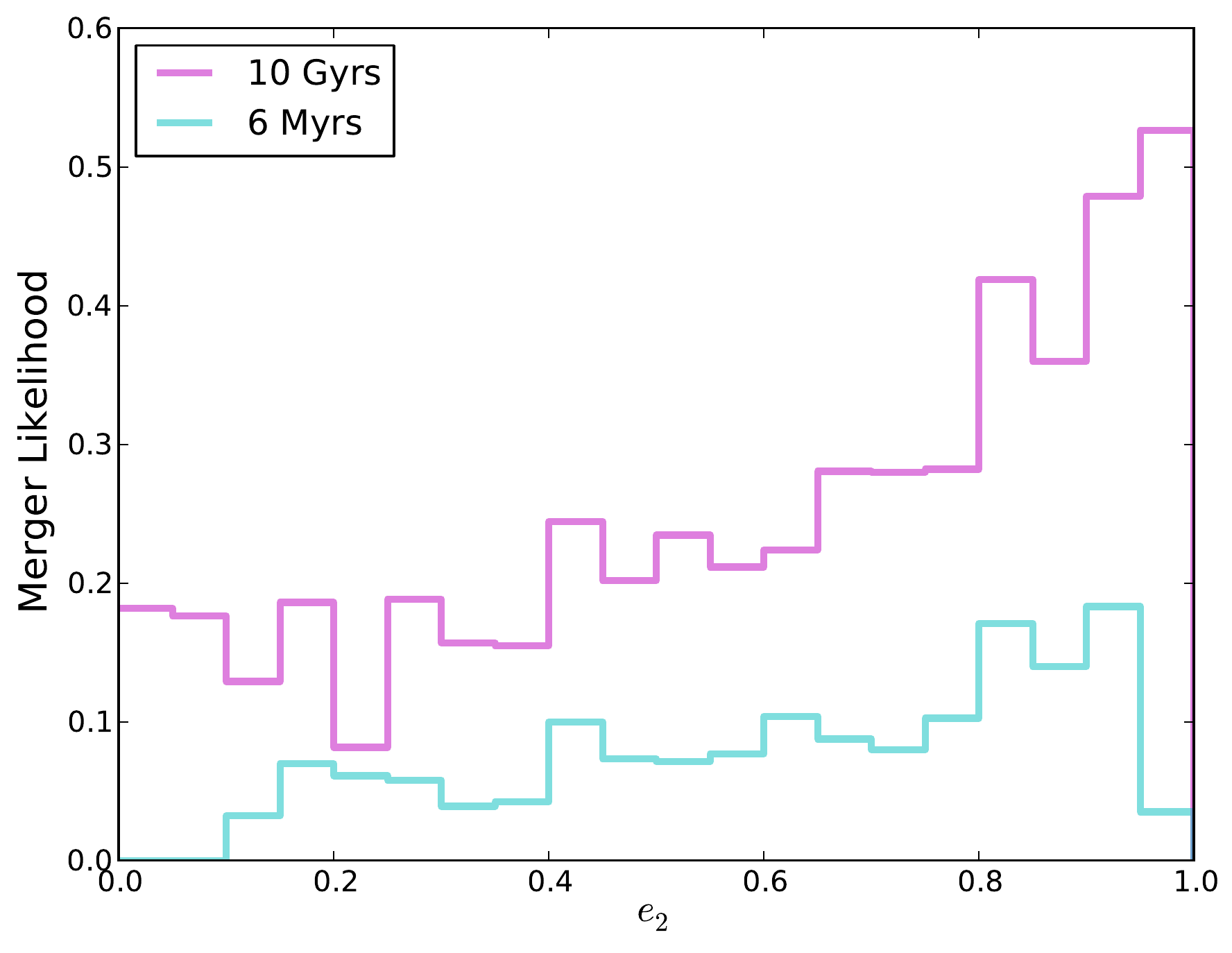}
\caption{{\bf Merger likelihood as a function of eccentricity.} Plotted are the fractions of merged systems as a function of the initial binary's orbital eccentricity around the MBH, shown after $6$~Myrs (cyan curve) and $10$~Gyrs (magenta curve). Note that the curves are normalized with the initial $e_2$ distribution after rejection (see Figure \ref{fig:distributions}), thus showing the merger fraction for a flat $e_2$ distribution. Both curves show a clear preference for higher eccentricities for mergers to form, which makes intuitively sense since larger eccentricities around the MBH induces stronger EKL interactions. At late times, however, low eccentricities can also yield a significant merger fraction. This can be explained with tidal effects and the evaporation time. These low-eccentricity systems had a higher long-term stability against evaporation and their weaker EKL oscillations allowed the systems to become tidally captured instead of directly merging.} 
   \label{fig:fracVSe2}
\end{figure}

The binary orbital configuration around the MBH (referred to here as the outer orbit)
sets limits to the different outcomes of the inner orbit, and thus a promising observable is the outer orbit's period distribution.  As shown in Figure \ref{fig:incVSeps}, the merger outcome is very sensitive to $\epsilon$ and thus to the eccentricity and the outer orbit separation $a_2$. We note that the outer orbit separation from the MBH,  $a_2$,  does not change during the evolution, as a consequence  of the secular approximation. The outer orbit eccentricity does not change because the outer orbit carries most of the angular momentum in the system, and thus the changes onto $e_2$ are insignificant compared  to the angular momentum variation of the inner orbit. The $e_2$ distribution of all merged systems  is shown  in Figure \ref{fig:fracVSe2}. G2-like candidates, i.e., those binaries that merged in the last $6$~Myrs, are preferentially on eccentric orbits, with a long tail down to $e_2\sim 0.1$. As time goes by, stellar evolution merger products become an important component of the overall merger population and allow for smaller values in the $e_2$ distribution. However, highly eccentric outer orbits are still preferred for forming mergers. This is due to the stronger EKL oscillations for eccentric outer orbits that enhance the formation of tidally locked systems.

\begin{figure}
\hspace{-0.075\columnwidth}
\includegraphics[width=1.1\columnwidth ]{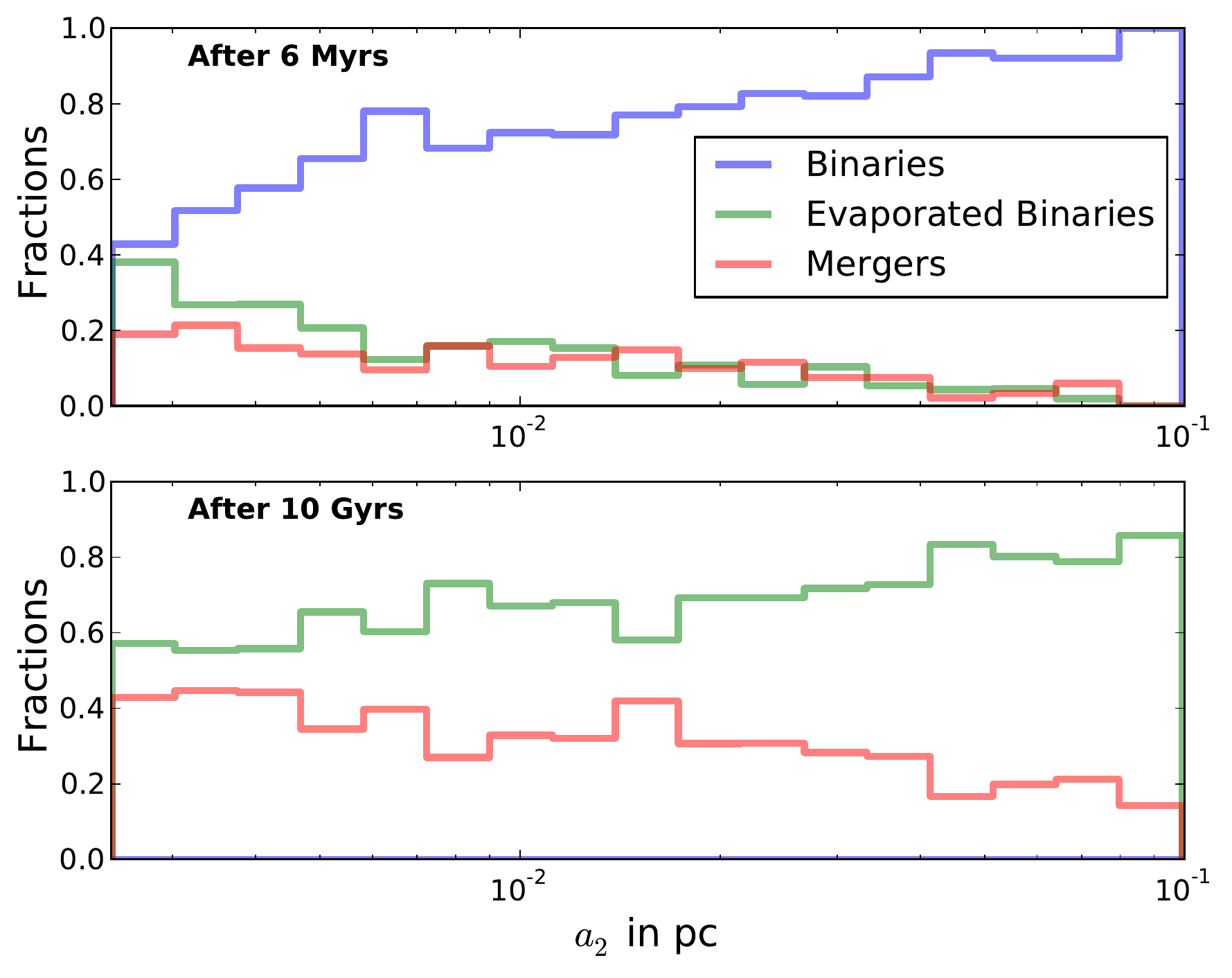}
\caption{{\bf Binary, evaporated, and merger fractions vs. $a_2$.} Plotted are the fraction of binaries, evaporated binaries, and merged binaries as a function of $a_2$, the semi-major axis of the binary's orbit around the MBH. Shown are the distributions at two timesteps, $6$~Myrs (upper panel) and $10$~Gyrs (lower panel). Note that the distributions of mergers and evaporated binaries show a preference for smaller $a_2$ after $6$~Myrs and a lack of mergers at the largest $a_2$ values, but this preference is strongly reduced after $10$~Gyrs. This makes intuitively sense since the binaries with smaller $a_2$ values will undergo stronger EKL evolution, thus they will merge faster and tidally lock faster, while they are also more likely to be evaporated due to a higher likelihood of scattering events. Binaries further away live longer, undergo weaker EKL oscillations, and have more time to become tidally locked, which increases the merger fraction at large $a_2$ values once the binary leaves the main sequence. After $10$~Gyrs all stars in our sample have been either evaporated or have undergone post main-sequence stellar evolution, so there will be no binaries of the original population left after this time. }
   \label{fig:fracVSa2}
\end{figure}

Another potential observable is the separation of merger products and binaries from the MBH. Again, we expect a strong transition between the two timescales considered throughout the paper. As can be seen in the top panel of Figure \ref{fig:fracVSa2}, G2-like candidates  have a long tail distribution with a preference to close separations. After $10$~Gyr the population of merger products is more uniformly distributed as a function of $a_2$, with a continued slight preference for smaller $a_2$ values (bottom panel).

{{ Finally we consider the validity of our secular approach in light of the extremely dense nature of the GC.
The double averaging method applied here will break when
the value of the inner orbit angular momentum goes to zero (i.e., extreme inner orbit eccentricity) on shorter timescale than the inner  orbital period \citep[e.g.,][]{Ivanov+05,Katz+12,Antognini+13,Antonini+14,Bode+14,Luo+16}. To test the validity of our method we use the criterion described by \citet{Katz+12} and \citet{Antonini+14} (for the inner orbit) , i.e.,
\begin{equation}\label{eq:Antonini_stab}
\sqrt{1-e_1} \lesssim \frac{15/2}{\sqrt{2}} \pi \frac{m_{\rm{MBH}}}{m_{\rm{binary}}}\left[\frac{a_1}{a_2(1-e_2)}\right]^3
\end{equation}
{ If this criterion is fulfilled, the secular approach will underestimate the strength of the dynamical evolution; however, only about $10\%$ of merged system and $10\%$ of all un-merged systems enter this regime during their EKL evolution. 
In most cases the expected consequences of violating the double averaging process, either for the inner or the outer orbit, are higher eccentricity excitations, which may also take place on slightly different timescales. {We conclude that we may be underestimating the rate of direct mergers and therefore also the total fraction of merged systems.} The timescales for these mergers are on the order of a few to tens of quadrupole timescales, roughly $10^5$ years. }
}}

\section{Discussion}\label{sec:dis}

We have studied the secular evolution of binary stellar systems in the GC while considering the octupole level of approximation for the hierarchical three-body problem. We include GR, tidal effects, and post main-sequence stellar evolution. The latter  includes mass loss, inflation of stellar radii, and magnetic braking for both stars.  
We predict a population of merged products that should exist in the GC.

 We  identify three distinct formation channels for merged systems.  As a system evolves, the EKL mechanism can cause large eccentricity excitations for the inner orbit.  If the eccentricity excitation happens on a relatively short timescale compared to the short range forces such as tides or GR, the binary members may cross each other's Roche limits (see left panels of Figure \ref{fig:largeplot}); we refer to this channel as {\bf direct mergers}. 
On the other hand, during close pericenter passages when tides are important, the orbital energy is dissipated and the orbit circularizes while the separation shrinks. In this case, as the radius of the more massive star starts to expand as it exits the main sequence, it overflows its Roche lobe and the system merges (see right panels of Figure \ref{fig:largeplot}). We refer to this channel as {\bf tidal mergers}. Even binaries that neither undergo strong eccentricity excitations nor become tidally locked can become mergers due to radial expansion of the more massive binary member, if the periapsis distance is smaller than the expanded star's new Roche limit (see central panels of Figure \ref{fig:largeplot}). We refer to this channel as {\bf radial mergers}.


Direct mergers happen only early in the systems' evolution, up to the first few million years after formation\footnote{Note that VRR is neglected in this calculation since direct mergers take place before VRR can have an effect. Furthermore, VRR acts to refill the high inclination parameter space that can lead to strong EKL evolution. By ignoring VRR effects we are effectively underestimating the direct merger rate, which makes our results a conservative estimate}. After roughly $6$~Myrs of evolution, $13 \%$ of all our binaries will have merged as direct mergers (see  Figures \ref{fig:timeVSmass} and  \ref{fig:fracVStime}). The merger rate then stalls and the evolution of the remaining binaries can continue on one of three possible routes:  (1) they can either evaporate without any further interesting interactions; (2) they can continue to undergo small amplitude eccentricity and inclination oscillations for a significant amount of time; or (3) they can undergo strong tidal interactions which circularize and shrink their inner orbits. After a few million years, as the massive stars start to expand in radius due post main sequence evolution, the merger rate begins to increase again as the surviving binaries become radial or tidal mergers. The total number of mergers increases over the next few Gyrs until the fraction of available binaries goes to zero due to evaporation and mergers (see plateau in Figure \ref{fig:fracVStime} for times $>10^9$~yrs). At this stage the total merger fraction is around $29 \%$, $35 \%$ of which are direct mergers from the first few million years of evolution, $10 \%$ of which are radial and $55 \%$ are tidal mergers.

However, for these merger fractions we assumed a minimum survival time of $10$~Gyrs for hard binaries, which gave many tidally locked binaries time to evolve past the main sequence and allowed them to merge during expansion. If we instead ignore the formation of hard binaries and continue to treat their evaporation time as we have for soft binaries, the total merger fraction is reduced to approximately $18 \%$ since most tidally locked binaries would evaporate before they expand. See Appendix \ref{App:AppendixA} for a justification of using the longer survival time, which yields more mergers.

If a star formation episode took place about $6$~Myrs ago we suggest that the direct merger population are G2-like candidates. As the overall evolution of these merged products is still unclear, we can only speculate{, based on observations of dusty binary mergers \citep[e.g.,][]{Tylenda+2011_1, Tylenda+2011_2, Tylenda+2013, Nicholls+2013} and the current state of common envelope evolution theory \citep[see for review][]{Ivanova+2013}, } that they will harbor extended gas and dust envelopes, which match G2 observations. Radial mergers would occur after $\sim 10$~Myrs until $\sim 200$~Myrs after star formation as the most massive stars begin to evolve past the main sequence and expand radially. This does not work for lower mass star binaries as these will evaporate before stellar evolution can occur. However, mergers can occur for smaller mass stars whose orbits have been tidally shrunken and circularized, extending their lifetime against evaporation. These tidal mergers begin to occur after a few tens of Myrs until a few Gyrs after star formation. There is considerable overlap in the occurrence times for radial and tidal mergers around the $100$~Myrs time mark, due to competing timescales for the onset of stellar evolution and tidal circularization of the inner orbit for stellar masses between $3$~and $10$~$M_\odot$. 
Assuming the earliest star formation episode occurred $10$~Gyrs ago, all of these old stellar binaries will have evaporated or merged by now. These merger products would have had time to cool down and relax and may look like young stars compared to the surrounding stars \citep[e.g.,][]{Antonini+2011} \citep[similar arguments were done for blue stragglers by][]{PF09,Naoz+14stars}.

Thus, we distinguish between the early merger population and the late population. These populations also show somewhat different orbital element distribution (see Figures \ref{fig:fracVSe2} and \ref{fig:fracVSa2}). Early, direct mergers show a slight preference for small separations and high orbital eccentricities around the MBH, which is less profound for late (radial and tidal) mergers. This is consistent with the EKL mechanism for which closer eccentric binary orbit  around the MBH may result in larger eccentricity excitations.  However, for radial and tidal mergers the dominating factor is the survival time of a system against evaporation, which is longer for larger semi-major axis values and smaller eccentricities.

We also find that the efficiency of tides has a negligible effect on the results. This is because the direct mergers are independent on the tides, and the radial and tidal mergers are more sensitive to the post main sequence radial expansion. 

{We conclude that there is a possible connection between the binary mergers described in this work and the formation of millisecond pulsars (MSP). \citet{Macquart+2015} have recently pointed out that the GC could harbor a relative overabundance of MSPs, although they argued that this would be due to the dense nature of the GC stellar environment. Binary mergers, however, could naturally produce such rapidly spinning pulsars due to the need of conserving the binaries' orbital angular momentum. }

We note that we do not speculate here on how the binaries arrived at their separations from the MBH. Instead we pose a simpler question of what will be the resulting dynamical evolution of a binary at a given distance from the MBH. It is unclear if these binaries could have formed there as the gravitational and tidal forces should prevent the formation of stars so close to the MBH \citep[see, for example,][]{AllenSanders1986, Morris1993, Ghez+2003, Alexander2005, Genzel+2010}. However \citet{LevinBelo2003} have claimed that {\it in situ} star formation close to the MBH would be possible in a dense gaseous disk \citep[see][for similar ideas]{Amaro+14,Chen+14}. That disk could have existed around the MBH in the past, formed either through gradual accumulation of infalling gas or tidal disruption of an infalling molecular cloud. However, it has been suggested that the presence of such a gaseous disk could lead to efficient merging of binaries, see \citet{Bartos+16}.  If the binaries arrived to their location via two body relaxation processes \citep[see][]{AntPer2012}, then the precession due to this process should also be taken into account.

The merger products formed through these mechanisms can undergo very interesting and potentially wildly differing further evolution. The direct merger scenario can lead to very violent mergers through collisions or grazing of the stars' outer envelope. Radial mergers will lead to Roche-lobe overflow with a comparatively wide, moderately eccentric stellar orbit, while tidal mergers will lead to common-envelope evolution of circularly and tightly orbiting stars. All these mechanisms will lead to highly complex and interesting mass-transfer and post main sequence evolutions. Regardless of the details of the processes mentioned above, these merger products represent a new and interesting class of objects that should be present in the GC.

\section*{Acknowledgements}

APS and SN acknowledge the partial support from the Sloan foundation. We thank Avi Loeb for insightful discussions and we also thank Melvyn Davis, Tal Alexander, and Fabio Antonini for useful discussions that took place during the 2016 Aspen winter conference on Dynamics and Accretion at the Galactic Center. SN and APS thank Tassos Fragos and Aaron Geller for assistance in the initial stages of code development. We thank the reviewer, Joseph Antognini, for his helpful comments and suggestions for improving this paper.
This work was supported in part by the National Science Foundation through grant AST-1412615 (awarded to AMG), as well as by the European Research Council under the European Union's Horizon 2020 Programme, ERC-2014-STG grant GalNUC 638435 (awarded to BK).




\bibliographystyle{mnras}
\bibliography{Kozai} 



\appendix
\section{Binary Evaporation Timescale}\label{App:AppendixA}

{We closely follow the work done by \citet{Binney+Tremaine1987} for the evaporation timescales of binaries at the GC, but allow different masses for the  binary companions and interacting stars.

The dynamical behavior of binaries in stellar groups and clusters can be generalized into two separate cases, i.e. {\bf soft} and {\bf hard} binaries. Given a binary with average orbital energy
\begin{equation}\label{eq:orbE}
\widetilde{E} =-\frac{Gm_{1}m_{2}}{2a_1}
\end{equation}
that travels through a cluster with velocity dispersion $\sigma$, the binary is soft if the energy of its center of mass motion relative to the cluster is larger than $\widetilde{E}$, i.e.
\begin{equation}\label{eq:soft}
\frac{2|\widetilde{E}|}{(m_{1}+m_{2}){\sigma}^2} < 1
\end{equation}
In general, soft binaries become softer through their interactions with other cluster stars until they evaporate, while hard binaries become harder until they merge or other forces become dominant \citep[see, for example,][]{Heggie1975,Hut1983,Hut+1983,Heggie+1996}. But, as has been pointed out by \citet{Hopman2009}, it is not obvious that this general rule holds true close to a MBH due to the more violent nature of the environment and the tidal influence of the MBH. However, binaries in the GC need to have extremely small inner orbital separations to qualify as hard binaries, on the order of $\sim 0.1 ~AU$ or less. At these separations the stars will feel tidal forces from their binary companions which can help to stabilize the systems against external influences.

The softness condition is usually strongly satisfied by the GC. Assuming a velocity dispersion of $\sigma = 280 ~km ~s^{-1} \sqrt{0.1~ pc/a_2} $ \citep{KocTrem2011}, the ratio of the energies is small for nearly all systems under consideration (smaller than $\sim 0.2$ even for two $100$ M{$_\odot$} binary companions on a $a_1 \sim 1~AU$ orbit with a MBH semi-major axis of $a_2 \sim 0.1~pc$, assuming an average cluster star mass of $1$ {$M_\odot$}). This means that the GC binaries will become wider over time through encounters with other stars in the cluster.

The widening or ``softening'' of soft binaries works through close encounters of field or cluster stars with one of the binary companions. Since the velocity dispersion is high compared to the orbital velocity of the binary companions, such an encounter will have a much stronger influence on the companion closer to the passing third star than on the farther companion. The change of internal energy can be written as
\begin{equation}\label{eq:deltaE}
\Delta \widetilde{E}=\frac{1}{2}\mu\Delta (V^{2})=\mu(\Delta {\bf v_{1}}({\bf v_{1}}-{\bf v_{2}})+ \frac{1}{2} (\Delta {\bf v_{1}})^{2})
\end{equation}
with $\mu = \frac{m_1 m_2}{m_1+m_2}$ and the ${\bf v}$ vectors describing the velocities of the binary companions. Following the derivation of \citet{Binney+Tremaine1987}, but allowing for different binary companion masses $m_1$ and $m_2$ and cluster star masses $m_3$, we derive an energy diffusion rate of
\begin{equation}\label{eq:diffusion}
\langle D(\Delta \widetilde{E}) \rangle = 16\sqrt{\frac{\pi}{3}}\frac{G^{2}\rho\ln \Lambda}{\sigma}\frac{m_{1}m_{2}m_{3}}{m_{1}+m_{2}}
\end{equation}
with a stellar mass density in the cluster of $\rho=1.35\times 10^6 ~M_\odot ~pc^{-3} {(a_2/0.25~pc)}^{-1.3}$ {\citep[see][]{Genzel+2010}}, and $\ln \Lambda$ being the Coulomb logarithm with $\Lambda = 15$.

Combining Equations \ref{eq:orbE} and \ref{eq:diffusion} we can now find the expected lifetime of soft binaries in the GC, their {\it evaporation time}:
\begin{equation}\label{eq:evap}
t_{EV} = \frac{\sqrt{3}\sigma}{32\sqrt{\pi} G\rho a_1\ln \Lambda}\frac{m_{1}+m_{2}}{m_{3}}
\end{equation}
This solution is equivalent to the result in \citet{Binney+Tremaine1987} except for the last factor that includes the masses of the binary companions ($m_1$ and $m_2$) and of the average cluster star ($m_3$, assumed to be $\sim 1$ ~$M_\odot$) \citep[see also][]{Hopman2009}. This result immediately shows that massive binaries have a longer expected lifetime than binaries that are assumed to be comparable to the cluster stars. Considering the stellar IMF referred to in Section \ref{sec:Num}, the lifetime could therefore be longer by up to two orders of magnitude, a significant difference for our simulations as this allows stellar evolution to become relevant. Furthermore, reducing $a_1$ through tidal interactions will also increase the evaporation time, giving tidally evolved and locked systems more time to reach post main sequence evolution. 

While the conditions of the GC strongly favor binaries to be soft, we do observe that some binaries can become hard over time due to tidal energy dissipation and shrinking of their inner orbits. If the radial separation becomes so small that a binary qualifies as a hard binary we generally assume that the binary survives for at least $10$~Gyrs. As pointed out above, hard binaries will in general become harder through encounters with other stars, not softer. However, they can be ionized by single strong encounters, with an ionization probability per unit time of:
\begin{equation}\label{eq:ion_hard}
B(\widetilde{E}) = \frac{8 \sqrt{\pi} G^2 m^3 \rho \sigma}{3^{3/2} |\widetilde{E}|^2} \left( 1+ \frac{m \sigma^2}{5 |\widetilde{E}|} \right)^{-1} \left[ 1+ \exp \left( \frac{|\widetilde{E}|}{m \sigma^2} \right) \right] ^{-1}
\end{equation}
Equation \ref{eq:ion_hard} is taken from \citet{Binney+Tremaine1987} for an example of a binary with equal masses encountering another equal mass star. This probability becomes very small (on the order of  $\sim10^{-10}$ per year or smaller) for the hard binaries in our sample, thus justifying our choice of a $10$~Gyr survival time.
}


\bsp	
\label{lastpage}
\end{document}